# Thermoelectric Algebra Made Simple for Thermoelectric Generator Module Performance Prediction under Constant Seebeck-Coefficient Approximation

Byungki Ryu (류병기),* Jaywan Chung (정재환), and SuDong Park (박수동)

Energy Conversion Research Center, Korea Electrotechnology Research Institute (KERI), Changwon-si, 51543, Republic of Korea

**Abstract**

While thermoelectric material performances can be estimated using the *ZT*, predicting the performance of thermoelectric generator modules (TGMs) is complex due to the non-linearity and non-locality of the thermoelectric differential equations. Here, we present a simplified thermoelectric algebra framework for predicting TGM performance within the Constant Seebeck-coefficient Approximation (CSA). First, we revisit the Constant Seebeck-coefficient Model (CSM) to transform the differential equations into exact algebraic equations for thermoelectric heat flux and conversion efficiency in terms of the load resistance ratio and relative Fourier heat flux. Next, we introduce the CSA, where the Thomson term is neglected, and the device parameters are assumed to be fixed. We define average thermoelectric properties and device parameters at the zero-current condition using a simple temperature integral. Finally, we derive approximate thermoelectric algebraic equations for voltage, resistance, heat flux, and conversion efficiency as functions of current. We numerically validate that the CSA formalism is superior to other single-parameter theories, such as peak-*ZT*, integral-*ZT*, and the generic engineering-ZT, in predicting efficiency. The relative standard error in optimal efficiency is less than 11% for average ZT values not exceeding 2. By combining CSM and CSA, TGM performance can be easily estimated without the need for calculus or solving differential equations. Therefore, the simplified thermoelectric algebra under the CSA framework has the potential to significantly enhance future TGM analysis and design, facilitating more efficient device-level research and development beyond the traditional focus on material properties.

* Correspondence: byungkiryu@keri.re.kr

# 1. Introduction

**Thermoelectric phenomena** enable the direct conversion between thermal and electrical energy.[1,2] Thermoelectric technology has been considered for various applications, including power generation, thermometers, solid-state refrigeration, and active cooling devices and modules.[3,4] This technology shows great potential for waste heat recovery and small-scale chip cooling and thermal management.[5] When a temperature difference is applied across a thermoelectric element or leg, a voltage difference is generated via the Seebeck effect.[1,6,7] Similarly, when an electric current flows, a thermal current is produced via the Peltier effect.[1,6,7]

**Thermoelectric efficiency** ($\eta$) of materials is often characterized by the dimensionless figure of merit $ZT = \frac{\alpha^2}{\rho\kappa}T$, where thermoelectric properties (TEPs) $\alpha$, $\rho$, and $\kappa$ are Seebeck coefficient, electrical resistivity, and thermal conductivity, and $T$ is absolute temperature.[1,4] Higher $ZT$ values indicate greater thermoelectric conversion efficiency from thermal to electrical energy. In 1957, Ioffe derived the optimal thermoelectric efficiency, $\eta_{\text{opt}}$, which represents the maximum efficiency under a given temperature difference between the hot-side temperature $T_h$ and cold side temperature $T_c$, for temperature-independent TEPs.[3,7] This is given by the following expression:

$$\eta \leq \eta_{\text{opt}} = \eta_{\text{opt}}^{(\text{CPM})} := \frac{\Delta T}{T_h} \frac{\sqrt{1 + ZT_{\text{mid}}} - 1}{\sqrt{1 + ZT_{\text{mid}}} + \frac{T_c}{T_h}} \tag{1}$$

where $\Delta T := T_h - T_c$ and $T_{\text{mid}} := \frac{T_h + T_c}{2}$. The ZT-based performance estimation has been highly effective in the discovery and optimization of thermoelectric materials, helping to identify materials with higher $ZT$ values over time:[8] Before 2000, the maximum peak $ZT$ value was only around 1, but recent advancements have led to reported $ZT$ values exceeding 3.0.[9]

**However, a significant efficiency gap exists between thermoelectric materials and modules.** While the record efficiency of a multistage thermoelectric generator module (TGM) is 13%, this is considerably lower than the theoretical efficiency limit of 17% for an ideal material, which corresponds to an average $ZT$ value of around 1.[8] Several factors contribute to this discrepancy, including the overestimation of efficiency based on peak $ZT$, sub-optimal material selection, parasitic electrical and thermal resistance,[8] as well as mechanical failure and crack formation in the TGM during the joining process.[10] Additionally, long-term atomic diffusion and material instability in the TGM presents significant challenges.[11–13]

**The poor representation power of $ZT$ can be one of the reasons for the efficiency gap between materials and modules.** While $ZT$ provides a good theoretical efficiency limit for a given material, it hardly reflects the interactions between materials inside the modules and with external components. For example, a TGM requires a thermal-electrical circuit for power generation, where the generated current flows through an external load resistance. Heat flows into the TGM from the external environment and is then rejected. As a result, TGM performance is inherently dependent on both the current and the thermal environment. However, there is no way to express power and heat input using $ZT$ alone. Therefore, to further understand and improve the design of TGMs, it is necessary to also consider the electrical and thermal circuits, along with device parameters such as voltage ($V$), resistance ($R$), and thermal conductance ($K$) under the current ($I$) and environmental temperature conditions ($T$).

**In the meantime, various average $ZT$ schemes have been developed to overcome the limit in predicting material ZT performance over a wide range of temperatures.** In 2022, Chung, Ryu, and Seo further generalized the optimal efficiency formula under the Constant-Seebeck coefficient Model (CSM),[14] where only Seebeck coefficient is temperature-independent ($\alpha(T) = \alpha_0$). Within the CSM, the figure of merit $ZT$ and the optimal efficiency formula are generalized as:

$$Z_{\text{CSM}} = \frac{\alpha_0^2}{\frac{1}{\Delta T}\int_{T_c}^{T_h} \rho(T)\,\kappa(T)\,dT}, \quad (2)$$

$$\eta_{\text{opt}}^{(\text{CSM})} := \frac{\Delta T}{T_h} \frac{\sqrt{1 + Z_{\text{CSM}} T_{\text{mid}}} - 1}{\sqrt{1 + Z_{\text{CSM}} T_{\text{mid}}} + \frac{T_c}{T_h}}. \qquad (3)$$

When all thermoelectric properties (TEPs) are temperature-dependent, thermoelectric efficiency involves three degrees of freedom (DOFs): in addition to the general $ZT$, there are two additional degrees of freedom, $\tau$ and $\beta$, which are also thermoelectric figures of merit.[7] Although the three-DOF theory reveals a fundamental relationship between material property curves and efficiency, overcoming the counterintuitive relation between efficiency and classical $ZT$ ($ZT$ paradox[15]), it lacks a *simple* description of TGM performance. Similarly, an engineering $ZT$, incorporating Joule and Thomson heat corrections, has been reported,[16,17] but with a more complex optimal efficiency formula. While multiparameter approaches can provide accurate efficiency predictions by incorporating additional terms for temperature-dependent properties, the resulting TGM performance equations are not easily accessible, as they require three parameters, adding complexity.

**In this work, we aim to bridge the gap between material and module performance descriptions** by leveraging the simplicity of the $ZT$-based single-parameter model and the three thermoelectric degrees of freedom model. We develop a simplified thermoelectric algebra framework within the Constant-Seebeck coefficient Approximation (CSA). Revisiting the Constant-Seebeck coefficient Model, we derive exact expressions for thermoelectric heat flux and conversion efficiency, assuming a constant Seebeck coefficient. By neglecting the Thomson term and assuming fixed device parameters, we define average thermoelectric properties using a simple temperature integral. Finally, we approximate thermoelectric algebraic equations for both electrical and thermal circuits of TGMs, demonstrating that the CSA formalism outperforms other single-parameter models in efficiency prediction.

## 2. Theoretical results

### 2.1. One-Dimensional Model of Thermoelectric Generator Module

We consider a single-stage TGM, as shown in **Figure 1**. The thermoelectric generator module can be simplified as a one-dimensional thermoelectric single leg. The P-N leg-pair module can easily be generalized to multiple legs. A single leg may consist of a series of materials, but for simplicity, we assume that the leg is homogeneous, meaning that the thermoelectric properties of the leg are functions of temperature:

$$\alpha = \alpha(T), \tag{4}$$

$$\rho = \rho(T), \tag{5}$$

$$\kappa = \kappa(T). \tag{6}$$

We also assume that the leg material is p-type, with holes as the majority carriers, meaning the Seebeck coefficient is non-negative. The leg's cross-sectional area is fixed to $A$, and the leg length is $L$, from $x = 0$ to $x = L$.

### 2.2. Boundary Conditions and Thermal and Electrical Circuits outside the TGM

The left end of the leg is in contact with the hot side, while the right end is in contact with the cold side. Therefore, the following temperature boundary conditions apply. Currently, we assume Dirichlet boundary conditions:

$$T(0) = T_h, \tag{7}$$

$$T(L) = T_c, \tag{8}$$

$$\Delta T := T_h - T_c > 0. \tag{9}$$

Since $\Delta T$ is positive, an electromotive force is generated inside the leg, from the hot side to the cold side, as holes are the majority carriers. In the case of an open-circuit condition, there is no current. The left and right ends are connected to the load resistance $R_L$, with the left end of the leg set as the ground. The voltage $V_L$ is generated at the right end of the leg. However, the voltage distribution inside the leg will be discussed in a later section of the paper. The voltage $V_L$ is directly applied to the load resistance outside the module, as it is electrically connected. This causes a current $I$ to flow through the load, allowing the TGM to generate electricity and act as a power source for the load.

$$V(0) = 0, \tag{10}$$

$$V(L) = V_L, \tag{11}$$

$$\Delta V \coloneqq V_L - 0 = IR_L \tag{12}$$

2.3. **Thermoelectric Physics inside the TGM: Thermoelectric Differential Equations**

We focus on the steady-state performance prediction for the TGM. The performance can be described using the following two differential equations for the electrical current density $J \coloneqq \frac{I}{A}$ and thermal current density $J^Q \coloneqq \frac{Q(x)}{A}$ where $Q(x)$ is the thermal current. In solids, the charge carriers can flow due to the gradient of the chemical potential ($\mu$) of a given charge carrier, even in the absence of an electric field $E$:[6]

$$J = \sigma\,(\varepsilon - \alpha \nabla T), \tag{13}$$

$$\varepsilon = E + \frac{1}{e}\nabla \mu, \tag{14}$$

where $\varepsilon$ is the electromotive force and $e$ is the electron charge. In this work, we focus on continuum-level theory for macroscopic phenomena, while the chemical potential relates to the microscopic quantities with holes and electrons. To boldly simplify, we reduce the current density expression using only the macroscopic quantities: electric field and temperature. We believe this simplification does not cause any issues in circuit level description.

$$J = J(x) = \sigma\,(E - \alpha \nabla T), \tag{15}$$

$$E = -\nabla V \tag{16}$$

Note that, the direction of the electric field is from right to left, while the current flows from left to right. This is because that the TGM functions as a power source int the thermoelectric electrical circuit.

Additionally, due to *current conservation* in steady state, the $J$ remains constant across the positions in this one-dimensional model.

Meanwhile, the current flows through both thermal conduction and the Peltier effect. When the current is zero, only thermal conduction occurs:

$$J^Q(x|J=0) = -\kappa \nabla T. \tag{17}$$

With the Peltier effect in play, the heat current density equation becomes:

$$J^Q(x) = -\kappa \nabla T + (\alpha T) J. \tag{18}$$

It is crucial to note that thermoelectric performance is fully revealed by examining the internal physics and temperature distribution described by the thermoelectric differential equations **(15-18)** for electrical and thermal current densities, in combination with the electrical and thermal circuit boundary conditions: **equations (7-12)**.

The charge density must be conserved, while thermal energy must balance the dissipation of electrical energy. By applying the continuity equations for electrical and thermal current densities:

$$\nabla \cdot J = 0, \tag{19}$$

$$\nabla \cdot J^Q = E \cdot J, \tag{20}$$

we derive the differential equation for temperature within the TGM:[7,18]

$$\nabla \cdot (\kappa \nabla T) - T \left(\frac{d\alpha}{dT}\right)(\nabla T \cdot J) + \rho J^2 = 0. \tag{21}$$

Here, the thermal diffusion term balances the local Thomson heat and Joule heat. Focusing on the one-dimensional case, the partial differential equation simplifies to:

$$\frac{d}{dx}\left(\kappa \frac{dT}{dx}\right) - T\left(\frac{d\alpha}{dT}\right)\left(\frac{dT}{dx}\right)J + \rho J^2 = 0. \tag{22}$$

### 2.4. Device parameters and normalization

The device parameters of a TGM can be calculated over the spatial coordinate $x$ inside the leg, not over $T$, as discussed in the reference.[7] To achieve this, one must first determine the temperature solution of **equation (22)** under a given current density condition. With this solution, the average thermoelectric properties can be introduced as:

$$\bar{\alpha} := \frac{V_o}{\Delta T} = \frac{1}{\Delta T}\int_0^L (-\alpha \nabla T)\, dx, \tag{23}$$

$$\bar{\rho} := \frac{A}{L}R = \frac{1}{L}\int_0^L \rho\, dx, \tag{24}$$

$$\frac{1}{\bar{\kappa}} := \frac{A}{L}\frac{1}{K} = \frac{1}{L}\int_0^L \frac{1}{\kappa}\, dx. \tag{25}$$

It has been reported that these parameters in **equations (23-25)** vary slowly with current.[7] Therefore, above parameters can be approximated to the values when current is zero. When current is zero, the average TEPs and $Z$ are simply written as:

$$\alpha_0 := \bar{\alpha}_{(I=0)} = \frac{\int_c^h \alpha(T)\, dT}{\Delta T}, \tag{26}$$

$$\kappa_0 := \bar{\kappa}_{(I=0)} = \frac{\int_c^h \kappa(T)\, dT}{\Delta T}, \tag{27}$$

$$\rho_0 := \frac{\int_c^h \rho(T)\kappa(T)\, dT}{\kappa_0 \Delta T}, \tag{28}$$

$$Z_0 := \frac{\alpha_0^2}{\rho_0 \kappa_0} = \frac{\left(\int_c^h \alpha(T)\, dT\right)^2}{\Delta T \left(\int_c^h \rho(T)\kappa(T)\right)}. \tag{29}$$

Note that, the $Z_0$ corresponds to the one-shot approximation of the *general figure of merit* in three thermoelectric degrees of freedom theory.[7] This shape is the same to the *effective ZT* by Ioffe and Borrego.[3,19,20] The device parameters for voltage $V_0$, resistance $R$ and thermal conductance $K$ for the open-circuit TGM are:

$$V_0 := \alpha_0 \Delta T, \tag{30}$$

$$R_0 := \rho_0 \frac{L}{A}, \tag{31}$$

$$K_0 := \kappa_0 \frac{A}{L}. \tag{32}$$

Using **equations (30-32)**, we define the reference current $I_{ref}$ and reference current density $J_{ref}$:

$$I_{ref} := \frac{V_0}{R_0} = \frac{\alpha_0 \Delta T}{R_0} = \frac{\alpha_0 \Delta T}{\rho_0 L} A, \tag{33}$$

$$J_{ref} := \frac{I_{ref}}{A}. \tag{34}$$

Using **equations (26-28)**, we normalize the Seebeck coefficient, electrical resistivity, thermal conductivity, current, and current density as:

$$\hat{\alpha}(T) := \frac{\alpha(T)}{\alpha_0}, \tag{35}$$

$$\hat{\rho}(T) := \frac{\rho(T)}{\rho_0}, \tag{36}$$

$$\hat{\kappa}(T) := \frac{\kappa(T)}{\kappa_0}, \tag{37}$$

$$i := \frac{I}{I_{ref}}, \tag{38}$$

$$j := \frac{J}{J_{ref}} = i. \tag{39}$$

When thermoelectric properties are slowly varying functions of $T$, the normalized property curves $\hat{\alpha}$, $\hat{\rho}$, $\hat{\kappa}$ are close to 1. The normalized current will range from 0 to approximately 1. Finally, we can normalize the thermoelectric differential equation using normalized variables for temperature and position:

$$\hat{T} := \frac{T}{T_h}, \tag{40}$$

$$\hat{x} := \frac{x}{L}, \tag{41}$$

$$\frac{d}{d\hat{x}}\left(\hat{\kappa}\frac{d\hat{T}}{d\hat{x}}\right) + [-Z_0 T_h]\left[\frac{\Delta T}{T_h}\right]\left(\frac{d\hat{\alpha}}{d\hat{T}}\right)\left(\frac{d\hat{T}}{d\hat{x}}\right)j + [Z_0 T_h]\left[\frac{\Delta T}{T_h}\right]^2 \hat{\rho} j^2 = 0. \tag{42}$$

Each term represents the normalized diffusion, the normalized Thomson heating, and the normalized Joule heating term. The second and third terms are scaled with $Z_0 \Delta T$. For $Z_0 \Delta T \ll 1$ or $i = j \ll 1$, the Thomson and Joule terms can be neglected. If these terms grow larger, all three terms become comparable. Also, note that the normalized differential **equation (42)** is independent of geometry, with respect to $L$ and $A$.

### 2.5. Constant Seebeck-Coefficient Approximation (CSA)

**Equation (34)** is physically exact in one-dimensional TGM models when there are no additional heat losses. Now, by applying the approximation of a slowly varying Seebeck coefficient, where $\left(\frac{d\hat{\alpha}}{d\hat{T}}\right)\left(\frac{d\hat{T}}{d\hat{x}}\right) \ll 1$, the second term can be neglected. Under this approximation, the thermoelectric differential equation simplifies further to:

$$\frac{d}{d\hat{x}}\left(\hat{\kappa}\frac{d\hat{T}}{d\hat{x}}\right) + [Z_0 T_h]\left[\frac{\Delta T}{T_h}\right]^2 \hat{\rho} j^2 = 0. \tag{43}$$

From the currently reported maximum peak $ZT$ values over explored materials and the best thermoelectric efficiency of materials explored so far,[8] we can estimate the order of the normalized Joule heating term. The TGM power and efficiency might be maximized around $j \approx \frac{1}{2}$, while $[Z_0 T_h] \sim 1$, $\frac{\Delta T}{T_h} \leq \frac{600}{900} = \frac{2}{3}$ and $\hat{\rho} \approx 1$. Therefore, we estimate the term $[Z_0 T_h] \times \left[\frac{\Delta T}{T_h}\right]^2 \times \hat{\rho} j^2$ to be approximately $1 \times \left(\frac{2}{3}\right)^2 \times \left(\frac{1}{2}\right)^2 = \frac{1}{9}$. This indicates that the change in the temperature distribution inside the leg may not be significant compared to that of zero-current condition when thermoelectric properties are slowly varying with temperature. When the Seebeck coefficient is temperature-independent (CSM case), **equation (43)** becomes exact.[14]

### 2.6. Derivation of Constant Property Model

When thermoelectric properties are all temperature-independent (CPM case), **equation (42)** simplifies as $\hat{\alpha} = \hat{\rho} = \hat{\kappa} = 1$. The temperature solution for CPM satisfies the followings:

$$\kappa \frac{d^2 T}{dx^2} + \rho J^2 = \kappa \frac{d^2 T}{dx^2} + \frac{I^2 R_0}{A \cdot L} = 0, \tag{44}$$

$$\frac{d^2 \hat{T}}{d\hat{x}^2} + [Z_0 T_h] \left[\frac{\Delta T}{T_h}\right]^2 j^2 = 0. \tag{45}$$

The solution for $T(x)$ is a polynomial:

$$T(x) = -\frac{\rho J^2}{2\kappa} x^2 + Bx + C. \tag{46}$$

Using the boundary conditions for the hot and cold sides, given by **equations (7) and (8)**, the solutions for $T(x)$ and $\frac{dT(x)}{dx}$ are:

$$T(x) = -\frac{I^2 R_0}{2K_0}\left(\frac{x}{L}\right)^2 + \left(\frac{I^2 R_0}{2K_0} - \Delta T\right)\left(\frac{x}{L}\right) + T_h, \tag{47}$$

$$\frac{dT(x)}{dx} = -\frac{\Delta T}{L} + \frac{I^2 R_0}{K_0 L}\left(\frac{1}{2} - \frac{x}{L}\right). \tag{48}$$

From the heat current **equation (18)**, the boundary heat currents at the hot and cold sides and the power $P$ are rewritten as:

$$Q_h^{(\text{CPM})} = A \cdot J_h^Q = -A\kappa \frac{dT(x=0)}{dx} + I\alpha T_h \tag{49}$$

$$= K\Delta T + I\alpha T_h - \frac{1}{2}I^2 R,$$

$$Q_c^{(\text{CPM})} = A \cdot J_c^Q = -A\kappa \frac{dT(x=L)}{dx} + I\alpha T_c \tag{50}$$

$$= K\Delta T + I\alpha T_c + \frac{1}{2}I^2 R,$$

$$P^{(\text{CPM})} = Q_h^{(\text{CPM})} - Q_c^{(\text{CPM})} = I(\alpha_0 \Delta T - IR_0). \tag{51}$$

Using **equations (10-12) and (15)**, integrating **equation (16)** yields the voltage relation between the outside and inside:

$$E = -\nabla V = \rho J + \alpha \nabla T, \qquad (52)$$

$$\int_0^L E \, dx = V(0) - V(L) = -V_L = -IR_L, \qquad (53)$$

$$\int_0^L (\rho J + \alpha \nabla T) dx = IR - \int_c^h \alpha \, dT, \qquad (54)$$

$$V_L = \alpha_0 \Delta T - IR. \qquad (55)$$

Since the load voltage is applied to the load resistance while current flows, the power delivered to the load resistance is derived as:

$$P_L := IV_L = I^2 R_L = I(\alpha_0 \Delta T - IR). \qquad (56)$$

Note that the power and voltage **equations (55) and (56)** always holds, regardless of the temperature dependency of TEPs. In CPM, $\alpha$ is $\alpha_0$ and therefore power is always given by $P^{(\text{CPM})} = P_L$, while the heat **equations (49) and (50)** are only valid for the CPM case.

## 2.7. Derivation of Constant Seebeck-Coefficient Model

In this section, we derive the performance equations for the CSM. While the optimal efficiency equation was originally derived by Chung et al., we present an alternative approach to achieve the same results.[14] In the CSM, the Thomson term is zero because the Seebeck coefficient is temperature-independent. Therefore, we have the following:

$$J_h^Q = J\alpha_0 T_h - \kappa_h T_h', \qquad (57)$$

$$J_c^Q = J\alpha_0 T_c - \kappa_c T_c', \qquad (58)$$

$$p := \frac{P}{A} = J_h^Q - J_c^Q, \qquad (59)$$

$$(\kappa T')' + \rho J^2 = 0, \qquad (60)$$

where we replace $'$ with $\frac{d}{dx}$. Let us introduce the relative Fourier heat flux $u$ and load resistance ratio $\gamma$:

$$u := -\frac{\kappa \nabla T}{J} = -\frac{\kappa T'}{J} \qquad (61)$$

$$\gamma := \frac{R_L}{R}, \qquad (62)$$

$$R_L = \gamma R, \qquad (63)$$

$$I = \frac{V}{R} \cdot \frac{1}{1+\gamma}. \qquad (64)$$

Now we have the followings:

$$\frac{J_h^Q}{J} = \alpha T_h - u_h, \qquad (65)$$

$$\frac{J_c^Q}{J} = \alpha T_c - u_c, \qquad (66)$$

$$\frac{p}{J} = \frac{J_h^Q}{J} - \frac{J_c^Q}{J} = \alpha \Delta T - (u_h - u_c), \qquad (67)$$

$$\frac{du}{dx} + \rho J = 0. \qquad (68)$$

Let us multiply $u$ on **equation (68)**:

$$u\left(\frac{du}{dx}\right) = -u\rho J. \qquad (69)$$

Then we integrate the left-hand side (LHS) and right-hand side (RHS) simultaneously. The results are as follows:

$$(LHS) := \int_0^L u\left(\frac{du}{dx}\right) dx = \frac{1}{2}(u_{x=L}^2 - u_{x=0}^2) = \frac{1}{2}(u_c^2 - u_h^2), \qquad (70)$$

$$(RHS) := \int_0^L -u\rho J\, dx = -\int_{T_h}^{T_c} \rho(T)\, \kappa(T) dT = \rho_0 \kappa_0. \qquad (71)$$

Finally, we obtain the $u^2$-conservation relation:

$$u_h^2 - u_c^2 = -2\rho_0 \kappa_0 = -2R_0 K_0. \qquad (72)$$

We may directly integrate **equation (68)** directly over the position $x$.

$$-\frac{du}{dx} = \rho J, \tag{73}$$

$$(LHS) := -\int_0^L \frac{du}{dx} dx = u_h - u_c, \tag{74}$$

$$(RHS) := \int_0^L \rho J \, dx = \bar{\rho} L J = \frac{\bar{\rho} L}{A} J A = RI. \tag{75}$$

By equating the LHS and RHS, we obtain the $u$-conservation relation between $u$, $J$, $I$, $R$, and $\gamma$.

$$u_h - u_c = I \times R = \frac{V_0}{R(1+\gamma)} \times R = \frac{V_0}{1+\gamma}. \tag{76}$$

Alternatively, **equation (76)** can be derived from **(67)** using **(65)** and **(66)**:

$$V_L = IR_L = \frac{p}{J} = \alpha \Delta T - (u_h - u_c), \tag{77}$$

$$(LHS) := \frac{I^2 R_L}{AJ} = IR_L = V_0 \frac{\gamma}{1+\gamma}, \tag{78}$$

$$(RHS) := V_0 - (u_h - u_c), \tag{79}$$

$$u_h - u_c = V_0 \left(1 - \frac{\gamma}{1+\gamma}\right) = \frac{V_0}{1+\gamma}. \tag{80}$$

Note that, although $J$ and $I$ are non-local parameters, the relative load resistance $\gamma$ is a good parameter for describing the relative Fourier heat current $u$. The two unknown parameters $u_h$ and $u_c$ can be solved from **equations (72) and (76)**.

$$X := \frac{V_0}{1+\gamma}, \quad Y := R_0 K_0, \tag{81}$$

$$u_h - u_c = X, \quad (u_h - u_c)(u_h + u_c) = -2Y, \tag{82}$$

$$u_h + u_c = -\frac{2Y}{X}, \tag{83}$$

$$u_h = \frac{X}{2} - \frac{Y}{X}, \tag{84}$$

$$u_c = -\frac{X}{2} - \frac{Y}{X}. \tag{85}$$

Using **equations (65) and (66), coupled with (84) and (85)**, we obtain:

$$\frac{J_h^Q}{J} = \alpha_0 T_h - u_h = \frac{Y}{X} + \alpha_0 T_h - \frac{X}{2}, \tag{86}$$

$$\frac{J_c^Q}{J} = \alpha_0 T_c - u_c = \frac{Y}{X} + \alpha_0 T_c + \frac{X}{2}. \tag{87}$$

Using device parameters, we may rewrite it as:

$$\frac{Q_h^{(CSM)}}{I} := \frac{J_h^Q}{J} = \frac{\int \rho \kappa \, dT}{\left(\frac{V_0}{1+\gamma}\right)} + \alpha_0 T_h - \frac{1}{2}\frac{V_0}{1+\gamma} \tag{88}$$

$$= \frac{I}{I} \times \left(\frac{R_0 K_0 \Delta T}{IR} + \alpha_0 T_h - \frac{1}{2}IR\right)$$

$$= \frac{1}{I}\left[(K_0 \Delta T)\left(\frac{R_0}{R}\right) + I\alpha_0 T_h - \frac{1}{2}(I^2 R_0)\left(\frac{R}{R_0}\right)\right].$$

## 2.8. Optimal efficiency for Constant Seebeck-Coefficient Model

Using **equations (78) and (88)**, we derive the efficiency formula as

$$\frac{P_L}{I} = \frac{P_L^{(CSM)}}{I} = V_L = IR_L = \alpha_0 \Delta T \left(\frac{\gamma}{1+\gamma}\right), \tag{89}$$

$$\frac{Q_h^{(CSM)}}{I} = \frac{J_h^Q}{J} = \frac{\int \rho \kappa \, dT}{\left(\frac{\alpha_0 \Delta T}{1+\gamma}\right)} + \alpha_0 T_h - \frac{1}{2}\left(\frac{\alpha_0 \Delta T}{1+\gamma}\right), \tag{90}$$

$$\frac{1}{\eta^{(CSM)}(I)} = \frac{Q_h^{(CSM)}}{P_L^{(CSM)}} = \frac{1}{\gamma} \cdot \left(\frac{(1+\gamma)^2}{Z_0 \Delta T} + \frac{(1+\gamma)}{\eta_{Carnot}} - \frac{1}{2}\right), \tag{91}$$

where $\eta_C$ is the Carnot efficiency, $\eta_C := \frac{\Delta T}{T_h}$. Note that the efficiency can be expressed as a function of $\gamma$. As the derivation to find the optimal efficiency is missing in literatures, yet it is important, we fully derive the equations in detail. A simple derivation is possible using the inequality of arithmetic and geometric means without the need of Calculus:

$$A\gamma + B\gamma^{-1} \geq 2\sqrt{AB}. \tag{92}$$

The equality holds when:

$$A\gamma_{opt} = B\frac{1}{\gamma_{opt}}. \tag{93}$$

We may obtain the following:

$$\frac{1}{\eta^{(CSM)}(I)} = \left(\left(\frac{1}{Z_0\Delta T}\right)\gamma + \left(\frac{2}{Z_0\Delta T} + \frac{1}{\eta_C}\right)\gamma^0 + \left(\frac{1}{Z\Delta T} + \frac{1}{\eta_C} - \frac{1}{2}\right)\gamma^{-1}\right), \tag{94}$$

$$\frac{1}{\eta^{(CSM)}(I)} \geq \frac{1}{\eta^{(CSM)}_{opt}} := \left(\frac{2}{Z_0\Delta T} + \frac{T_h}{\Delta T}\right) + 2\sqrt{\left(\frac{1}{Z_0\Delta T}\right)\left(\frac{1}{Z_0\Delta T} + \frac{T_h}{\Delta T} - \frac{1}{2}\right)}, \tag{95}$$

$$\frac{1}{\eta^{(CSM)}_{opt}} = \left(\frac{1}{Z_0\Delta T}\right)(2 + Z_0 T_h) + 2\sqrt{\left(\frac{1}{Z_0\Delta T}\right)^2\left(1 + Z_0\left(T_h - \frac{\Delta T}{2}\right)\right)} \tag{96}$$

$$= \frac{T_h}{\Delta T} + \left(\frac{2}{Z_0\Delta T}\right)\left[\sqrt{1 + Z_0 T_{mid}} + 1\right],$$

$$\frac{1}{\eta^{(CSM)}_{opt}}\frac{1}{\sqrt{\bullet} - 1} = \left(\frac{T_h}{\Delta T}\right)(\sqrt{\bullet} - 1) + \left(\frac{2}{Z_0\Delta T}\right)(1 + Z_0 T_{mid} - 1) \tag{97}$$

$$= \left(\frac{T_h}{\Delta T}\right)\sqrt{\bullet} - \frac{T_h}{\Delta T} + \frac{2T_{mid}}{\Delta T}$$

$$= \left(\frac{T_h}{\Delta T}\right)\sqrt{\bullet} + \frac{-T_h + T_h + T_c}{\Delta T}$$

$$= \left(\frac{T_h}{\Delta T}\right)\left(\sqrt{\bullet} + \frac{T_c}{T_h}\right),$$

where the square root here is defined as $\sqrt{\bullet} := \sqrt{1 + Z_0 T_{mid}}$ for simplicity in the derivations. Finally, the optimal efficiency for the CSM is derived as:

$$\eta^{(CSM)}(\gamma) \leq \eta^{(CSM)}_{opt} = \frac{\Delta T}{T_h} \cdot \frac{\sqrt{1 + Z_0 \cdot T_{mid}} - 1}{\sqrt{1 + Z_0 \cdot T_{mid}} + \frac{T_c}{T_h}}, \tag{98}$$

where the equality holds under the following condition:

$$\left(\frac{1}{Z_0\Delta T}\right)\gamma_{opt} = \left(\frac{1}{Z\Delta T} + \frac{1}{\eta_C} - \frac{1}{2}\right) \cdot \frac{1}{\gamma_{opt}}, \tag{99}$$

$$\gamma_{\mathrm{opt}} = \frac{R_L}{R} = \sqrt{1 + Z_0 T_{\mathrm{mid}}}. \tag{100}$$

Note that $Z_0 = \frac{\alpha_0^2}{\rho_0 \kappa_0} = \frac{(\int \alpha dT)^2}{\Delta T (\int \rho \kappa \, dT)}$ in the **equation (100)** is similar to the $Z_{\mathrm{CSM}} = \frac{\alpha^2}{\Delta T (\int \rho \kappa \, dT)}$ in the **equation (2)**.

### 2.9. CSM comparison to the Three Thermoelectric Degrees of Freedom (DOFs) Theory

In the one-dimensional case, this differential **equation (22)** can be transformed into an integral equation for temperature, as reported in the literature.[7] Furthermore, without any assumptions, the equation can be transformed into algebraic equations for thermal current:[7]

$$Q_h^{(\mathrm{DOFs})}(I) = K\Delta T + I\bar{\alpha}(T_h - \tau\Delta T) - \frac{1}{2}I^2 R(1+\beta), \tag{101}$$

$$Q_c^{(\mathrm{DOFs})}(I) = K\Delta T + I\bar{\alpha}(T_c - \tau\Delta T) + \frac{1}{2}I^2 R(1-\beta), \tag{102}$$

$$P^{(\mathrm{DOFs})}(I) := Q_h - Q_c = I\bar{\alpha}\Delta T - I^2 R = I(\bar{\alpha}\Delta T - IR), \tag{103}$$

where $\tau$ and $\beta$ are related to the non-zero Thomson heating and asymmetric Joule heating inside the leg. For constant properties, it reduces to the CPM **equations (49-51)**.

When Thomson heating is zero ($\tau = 0$), $\beta$ still remains in addition to the general $Z_{\mathrm{gen}} = \frac{\bar{\alpha}^2}{RK}$ in the performance equations, while CSM indicates that there should be only one DOF in the CSM theory ($Z_0 = Z_{\mathrm{CSM}}$).[14] This can be understood as the explicit heat conduction term and the explicit half Joule heating term balancing each other, causing the effect of $\beta$ to vanish in the CSM. This might imply the following relation for all current conditions:

$$Q_h^{(\mathrm{CSM})}(I) = (K_0 \Delta T)\left(\frac{R_0}{R}\right) + I\alpha_0 T_h - \frac{1}{2}(I^2 R_0)\left(\frac{R}{R_0}\right), \tag{104}$$

$$Q_h^{(\mathrm{DOFs},\tau=0)}(I) = K\Delta T + I\bar{\alpha}T_h - \frac{1}{2}I^2 R(1+\beta), \tag{105}$$

$$Q_h^{(\mathrm{CSM})} = Q_h^{(\mathrm{DOFs},\tau=0)}, \tag{106}$$

$$R_0 K_0 \Delta T - \frac{1}{2} I^2 R^2 = RK \Delta T - \frac{1}{2} I^2 R^2 (1+\beta), \tag{107}$$

$$(RK - R_0 K_0)\Delta T - \frac{1}{2}(IR)^2 \beta = 0. \tag{108}$$

From numerical calculations, we validate that **equation (108)** holds for constant Seebeck coefficient.

### 2.10. Definition of Constant Seebeck-coefficient Approximation

Recall **equations (24) and (42)**: $R$ is not constant if $\rho(T)$ varies with temperature. But, if $\frac{\partial R}{\partial I} \approx 0$ and Thomson term is negligible, we may approximate it to zero current resistance $R_0$.

$$\left(\frac{d\hat{\alpha}}{d\hat{T}}\right)\left(\frac{d\hat{T}}{d\hat{x}}\right) j < \frac{1}{Z_0 \Delta T}, \quad \left(\frac{R_0}{R}\right) \approx 1, \tag{109}$$

$$Q_h^{(\text{CSA})}(I) := K_0 \Delta T + I\left(\frac{V_0}{\Delta T}\right) T_h - \frac{1}{2} I^2 R_0 \approx Q_h^{(\text{CSM})}(I). \tag{110}$$

Under the condition of **equation (109)**, the heat flux for the CSM can be approximated by **equation (110)**. We name this approximation framework as *Constant Seebeck-coefficient Approximation*.

### 2.11. Thermoelectric Algebra within Constant Seebeck-coefficient Approximation

Finally, under the CSA, we develop the thermoelectric algebra framework using the following thermoelectric algebraic equations:

$$V_L(I) = \int_0^L (-\alpha \nabla T)\, dx - I \int_0^L \rho(T)\, dx \approx V_L^{(\text{CSA})} := V_0 - IR_0, \tag{111}$$

$$P_L(I) = IV_L \approx P_L^{(\text{CSA})} := I(V_0 - IR_0), \tag{112}$$

$$Q_h(I) = -A\kappa \frac{dT}{dx} + I\alpha(T_h)T_h \approx Q_h^{(\text{CSA})} := K_0 \Delta T + I\left(\frac{V_0}{\Delta T}\right)T_h - \frac{1}{2}I^2 R_0 \tag{113}$$

$$\eta^{(\text{exact})} = \frac{P_L}{Q_h} \approx \eta^{(\text{CSA})} = \frac{P_L^{(\text{CSA})}}{Q_h^{(\text{CSA})}} \tag{114}$$

In the meantime, the optimal maximum efficiency can be found by maximizing the $\eta_{opt}^{(CSA)}$. Observed that $\eta^{(CSA)}(\gamma) = \eta^{(CSM)}(\gamma)$. Therefore $\eta_{opt}^{(CSA)}$ is the same to $\eta_{opt}^{(CSM)}$. We further simplify the **equations (111-114)** by introducing the normalized current $i = \frac{I}{I_{ref}}$ and the load resistance ratio $\gamma = \frac{R_L}{R}$, which is an effective approach for analyzing the experimental characterization curves of TGMs. **Table 1** summarizes the thermoelectric algebraic framework under the Constant Seebeck-coefficient approximation.

### 2.12. Theory extension toward P-N leg-pair TGM performance equations

While this work is based on the single-leg device performance, it can be easily extended to P-N leg-pair TGM performance analysis, as reported in the reference.[7] When considering the electrical and thermal circuits, the TGM connects P and N legs electrically in series and thermally in parallel, while the current and thermal boundary conditions are conserved. Therefore, the total voltage $V_L^{(pn)}(I)$ and thermal current should be summed. Since the current direction is opposite in N-type leg, we have

$$V_L^{(pn)}(I) := V_0^{(pn)} - IR_0^{(pn)} = V_L^{(p)}(I) - V_L^{(n)}(-I) \quad (115)$$

$$Q_h^{(pn)}(I) := K_0^{(pn)}\Delta T + I\alpha_0^{(pn)}T_h - \frac{1}{2}I^2 R_0^{(pn)} = Q_h^{(p)}(I) + Q_h^{(n)}(-I) \quad (116)$$

The above equations should hold for each current. Therefore, the CSA P-N leg-pair average TEPs and device parameters should be written as:

$$\alpha_0^{(pn)} := \frac{1}{\Delta T} V_0^{(pn)} = \alpha_0^{(p)} - \alpha_0^{(n)} \quad (117)$$

$$\rho_0^{(pn)} := \frac{A}{L} R_0^{(pn)} = \rho_0^{(p)} + \rho_0^{(n)} \quad (118)$$

$$\kappa_0^{(pn)} := \frac{L}{A} K_0^{(pn)} = \kappa_0^{(p)} + \kappa_0^{(n)} \quad (119)$$

Using these, we can apply the thermoelectric framework and algebraic equations from **Table 1.**

# 3. Numerical validation

## 3.1. Numerical Validation of Constant Seebeck-coefficient Approximation

To validate the CSA theory for predicting TGM performance accuracy, we compute the numerical optimal efficiencies of arbitrary thermoelectric property curves, with the working temperature ranging from 300 K to 900 K. For simplicity, we assume a linear TEP curve, such as $f(T) = (f_H - \alpha_C) \times \frac{(T-600K)}{600K} + f_0$, where $f(T)$ is one of $\alpha(T)$, $\rho(T)$, or $\kappa(T)$, $f_H \coloneqq f(T_h)$, $f_C \coloneqq f(T_c)$, and $f_0 \coloneqq \frac{f_H + f_C}{2}$. For $\rho(T)$ and $\kappa(T)$, $\frac{\rho_H - \rho_C}{2\rho_0}$ and $\frac{\kappa_H - \kappa_C}{2\rho_0}$ range from -0.8 to 0.8 with $\rho_0 = \kappa_0 = 1$. For $\alpha(T)$, $\frac{\alpha_H - \alpha_C}{2\alpha_0}$ ranges from -1 to +1 where $\alpha_0 = \sqrt{\frac{[Z_0 T_{\text{mid}}]}{T_{\text{mid}}} \times \rho_0 \kappa_0}$ with $\rho_0 = \kappa_0 = 1$. For $[Z_0 T_{\text{mid}}]$, we use values of 0.5, 1, 2, and 3. So we consider a very large TEP curve space, with a total of 5,324 material TEP curve sets: 4 average CSA $ZT$ values, $11 \times 11 \times 11$ uniform grid for $\alpha(T)$, $\rho(T)$, and $\kappa(T)$ linear-curve spaces, respectively. Of these, 5,295 TEP sets are well simulated.

**Figure 2** shows the numerical optimal efficiencies calculated using the integral formalism reported previously.[7,8] Note that when the slope of the Seebeck coefficient is zero, corresponding to the CSM case, the efficiency is simply determined by the $Z_0 T_{\text{mid}}$ value. When the Seebeck coefficient varies with temperature, the efficiency fluctuates around the CSM model efficiency. When the Seebeck coefficient variation is relatively small, the efficiencies are close to the efficiency corresponding to the given $Z_0 T_{\text{mid}}$ value. As the ZT or the slope of the Seebeck coefficient increases, the efficiency variation becomes larger. When the Seebeck coefficient slope is negative, the efficiency can exceed the CSM efficiency. This is due to the second figure of merit, $\tau$, from the three thermoelectric degrees of freedom.[7]

Next, we compare the formula-based optimal efficiencies from single- and multi-parameter models to the numerical values (see **Figures 3 and Table 2**). The relative error in efficiency is defined as the difference between the optimal efficiency formula and numerical optimal efficiency, expressed as a percentage of the numerical efficiency. The box and violin plots show the distribution of the relative error values. For optimal efficiency formula, we use the general maximum efficiency formula ($\eta_{max}^{gen}$) derived previously.[7] Also see the following:

$$\eta_{max}^{gen} = \frac{\Delta T}{T_h'} \cdot \frac{\sqrt{1 + Z_{gen} T_{mid}'} - 1}{\sqrt{1 + Z_{gen} T_{mid}'} + \frac{T_c'}{T_h'}}, \tag{120}$$

$$Z_{gen} = \frac{\left(\frac{V}{\Delta T}\right)^2}{RK} = \frac{\bar{\alpha}^2}{\bar{\rho}\bar{\kappa}}, \tag{121}$$

$$T_h' = T_h - \tau \Delta T, \quad T_c' = T_c - (\tau + \beta)\Delta T, \quad T_{mid}' = \frac{(T_h' + T_c')}{2}. \tag{122}$$

For the single-parameter theory, we replace $Z_{gen} T_{mid}'$ with each ZT value while $\tau = \beta = 0$. We consider the following single parameter models: $[ZT]_{max}$ as the peak $ZT$ model, $[ZT]_{int} = \frac{1}{\Delta T}\int_c^h ZT\, dT$ as the integral ZT model, $Z_{int} T_{mid} = \left(\frac{1}{\Delta T}\int_c^h Z\, dT\right) \cdot T_{mid}$ as the integral Z model, $Z_{eng} T_{mid} = \left(\frac{\left(\int_c^h \alpha\, dT\right)^2}{\left(\int_c^h \rho\, dT\right)\left(\int_c^h \alpha\kappa\, dT\right)}\right) \cdot T_{mid}$ as the single-parameter engineering Z model, using the CSA optimal efficiency formula. Additionally, $Z_0 T_{mid} = \left(\frac{1}{\Delta T} \cdot \frac{\left(\int_c^h \alpha\, dT\right)^2}{\int_c^h \rho\kappa\, dT}\right) T_{mid}$ represents the Constant Seebeck-coefficient Approximation. We also consider the multiparameter models. We also calculate the efficiency using the generic engineering ZT formula with corrected Thomson and Joule weight parameters, labelled as **Generic-Eng-ZT**. Finally, the results are compared to the three thermoelectric degrees of freedom model with exact $Z$, $\tau$, and $\beta$ values. To examine the effect of the number of degrees of freedom, we consider three distinct cases: ($Z00$) for the single-parameter model with exact $Z$ and $\tau = \beta = 0$; ($Z\tau 0$) for the single-parameter model with exact $Z$, exact $\tau$, and $\beta = 0$; and ($Z\tau\beta$) for the three-parameter model having exact $Z$, exact $\tau$, and exact $\beta$.

While $[ZT]_{max}$, $[ZT]_{int} = \frac{1}{\Delta T}$, and $Z_{int}T_{mid}$ methods overestimate the optimal efficiency, the CSA $Z_0T_{mid}$ model show the most accurate performance prediction, even compared to the exact Z model of (Z00) and the generic engineering ZT formula. However, this model is not superior to the general efficiency theory using the three thermoelectric degrees of freedom model because the CSA does not account for the temperature dependency in thermoelectric properties.

The CSA-based optimal efficiency prediction is very accurate among the one parameter models. This high accuracy in efficiency prediction was already recognized independently by Sherman[20] and Ponnusamy,[21] who compared the numerical efficiency and formula-based optimal efficiency. However, previous approaches lack sufficient data for numerical validation. Sherman considered three examples,[20] Ponnusamy considered six materials,[21] and Kim considered four material cases,[16] all of which are much smaller than our simulation size of approximately 5,000 TEP sets. In Ponnusamy's results, In Ponnusamy's results, it was stated that the exact constant property model using the exact $V$, $R$, and $K$ are the most accurate. However, in our validation over a large dataset, we find that the CSA with temperature averaging is superior to the CPM with spatial averaging for $R$ and $K$. The accuracy is very similar between these approaches, but Ponnusamy's method requires the calculation of $R$ and $K$, which in turn requires information about the temperature distribution at an optimal efficiency current condition. This means that efficiencies must be computed for all current points to determine the optimal efficiency and corresponding current. Though it is highly accurate, calculating all numerical efficiencies is impractical. In contrast, our CSA approach does not require any temperature profile information, as only the TEPs are integrated over temperature. Additionally, for the CSA framework, a single-shot optimal formula calculation is sufficient to determine the optimal efficiency.

**Figure 4** shows the variation of device parameters $R$ and $K$ with respect to the open-circuit values for the TGM under the operational condition of $T_h = 900\ K$ and $T_c = 300\ K$ for the material having linearly-dependent TEPs. The variation of parameters compared to the exact values are case-independent. But, we see some converged behavior when the CSA $Z_0 T_{\text{mid}}$ is not high or Seebeck coefficient variation is relatively small. As shown in **Figure 4a**, When $Z_0 T_{\text{mid}} = 1$ with $\frac{\Delta \alpha}{2\alpha_0} = 0.4$, $\frac{\Delta \rho}{2\rho_0} = -0.32$, $\frac{\Delta \kappa}{2\kappa_0} = -0.32$, although the TEP variation is quite large, the electrical and thermal resistance vary within 10%. Due to the relatively small CSM $Z_0 T_{\text{mid}}$, the heat current is highly accurately described. When the TEP variations become smaller, the resistance variations become decrease, indicating that the device performance is quite accurate when the CSA is hold: see **Figure 4b**. However, when the Seebeck coefficient variation and ZT value is not small, device parameters change with current and thereby the CSM heat current is highly different from the exact heat current, causing non-negligible errors: see **Figure 4c**. In this case, temperature-dependent TEPs should be treated with three thermoelectric degrees of freedom will be crucial to accurately describe the TGM performances.

## 4. Limitation

This work is valid for one-dimensional steady-state cases, homogeneous leg materials, Dirichlet thermal boundary conditions, slowly varying TEPs, and a relatively small current range. Otherwise, it may lead to errors. Additionally, we do not consider any contact or interfacial resistances. Heat loss due to radiation and convection is also not included. For large variations in TEPs, the algebraic equations may become inaccurate.

## 5. Summary

In summary, we provide the thermoelectric algebraic equations for predicting thermoelectric power generator module performances. Beginning with the thermoelectric differential equations, the relative Fourier heat flux equations at boundaries are solved using the load resistance ratio terminology. By applying the Constant Seebeck-coefficient Approximation, we define the approximate average thermoelectric properties and device parameters. This leads to a simplified thermoelectric equation framework for voltage, power, heat current, and efficiency as algebraic equations, eliminating the need for Calculus. Numerical analysis demonstrates that the thermoelectric algebraic formalism under the CSA is highly accurate for thermoelectric properties that vary slowly with temperature. Because the CSA framework requires no calculus, it effectively simplifies the characterization of thermoelectric generator modules. Furthermore, throughout the theory development, we provide a comprehensive description of how to derive Ioffe's figure of merit and the optimal efficiency formula, known as the Constant Property Model, as well as Constant Seebeck-coefficient Model. These derivations serve not only for educational purposes but also facilitate future theoretical advancements. The developed thermoelectric algebra is designed to be simple and easily applicable to thermoelectric generator module design and analysis, providing valuable insights for advanced device design and optimization that extends beyond material research.

# Acknowledgements

This work was supported by the Primary Research Program of KERI through the National Research Council of Science & Technology (NST) funded by the Ministry of Science and ICT (MSIT) (No. 24A01019), and the Korea Institute of Energy Technology Evaluation and Planning (KETEP) grant funded by the Ministry of Trade, Industry and Energy (MOTIE) (grant no. 2021202080023D)

# Conflict of Interest Statement

The authors have no conflicts to disclose.

# Author Contributions

**Byungki Ryu:** Conceptualization; Writing/Original Draft Preparation; Writing/Review & Editing.

**Jaywan Chung:** Conceptualization; Software; Writing/Review & Editing.

**SuDong Park:** Conceptualization; Funding Acquisition; Writing/Review & Editing.

# Data Availability Statement

The data that support the findings of this study are available from the corresponding author upon reasonable request.

# Tables

**Table 1.** Summary of thermoelectric framework under the Constant Seebeck-coefficient Approximation, including average thermoelectric properties, device parameters, electrical and thermal performance equations, and efficiency equations. For simplicity, we omit the notation of (CSA) in the following algebraic equations:

| Thermal working condition |
|---|
| $$\Delta T = T_h - T_c, \qquad T_{\text{mid}} = \frac{T_h + T_c}{2}$$ |
| **Average thermoelectric properties under CSA** |
| $$\alpha_0 = \frac{1}{\Delta T}\int_c^h \alpha(T)\,dT, \quad \kappa_0 = \frac{1}{\Delta T}\int_c^h \kappa(T)\,dT, \quad \rho_0 = \frac{1}{\kappa_0 \Delta T}\int_c^h \rho(T)\cdot\kappa(T)\,dT$$ |
| **Device parameters under CSA** |
| $$V_0 = \alpha_0 \Delta T, \quad K_0 = \kappa_0 \frac{A}{L}, \quad R_0 = \rho_0 \frac{L}{A}, \quad I_{\text{ref}} = \frac{V_0}{R_0}, \quad \gamma = \frac{R_L}{R_0}$$ |
| **Electrical Performance Equations under CSA** |
| $$I = I_{\text{ref}} i, \quad i = \frac{1}{1+\gamma},$$ $$V_L = V_0 - IR_0 = V_0(1-i) = \alpha_0 \Delta T \cdot \frac{\gamma}{1+\gamma}$$ $$P_L = IV_L = (I_{\text{ref}} V_0)\cdot i \cdot (1-i) = \left(\frac{\alpha_0^2 \Delta T^2}{R_0}\right)\cdot \frac{\gamma}{(1+\gamma)^2}$$ $$\frac{P_L}{K_0 \Delta T} = (Z_0 \Delta T)\cdot i \cdot (1-i) = (Z_0 \Delta T)\cdot \frac{\gamma}{(1+\gamma)^2}$$ $$\widehat{P_L} := \frac{P_L}{(K_0 \Delta T)(Z_0 \Delta T)} = i(1-i)$$ |
| **Thermal performance equations under CSA** |
| $$Q_h = K_0 \Delta T + I\alpha_0 T_h - \frac{1}{2}I^2 R_0$$ $$\frac{Q_h}{K_0 \Delta T} = 1 + (Z_0 T_h)\cdot i - \left(\frac{Z_0 \Delta T}{2}\right)\cdot i^2$$ $$\widehat{Q_h} := \frac{Q_h}{(K_0 \Delta T)(Z_0 \Delta T)} = \frac{1}{Z_0 \Delta T} + \frac{1}{\eta_C}\cdot i - \frac{1}{2}\cdot i^2$$ |
| **Efficiency equations under CSA** |
| $$Z_0 = \frac{\alpha_0^2}{\rho_0 \kappa_0}, \qquad \gamma_{\text{opt}} = \sqrt{1 + Z_0 T_{\text{mid}}}$$ $$\eta(I) := \frac{P_L(I)}{Q_h(I)} \approx \eta(\gamma) := \frac{P_L(\gamma)}{Q_h(\gamma)} \leq \eta_{\text{opt}}^{(\text{CSM})} = \frac{\Delta T}{T_h} \cdot \frac{\gamma_{\text{opt}} - 1}{\gamma_{\text{opt}} + \frac{T_c}{T_h}}$$ |

**Table 2. Root-mean square (RMS) of relative efficiency error for various efficiency formula models using various optimal efficiency formula with various single ZT parameter and multi-parameters.** Left columns for models and optimal efficiency formula. Right columns for selective ZT scales. With increasing ZT, efficiency RMS increases. The CSM efficiency prediction is the best among the single parameter theories. And its performance is also comparable to the (Z00).

| Model | | Optimal efficiency formula | $Z_0 T_{mid}$ values | | | |
|---|---|---|---|---|---|---|
| | | | 0.5 | 1 | 2 | 3 |
| Single ZT | ■ = $[ZT]_{max}$ | $\dfrac{\Delta T}{T_h} \cdot \dfrac{\sqrt{1+\blacksquare}-1}{\sqrt{1+\blacksquare}+\dfrac{T_c}{T_h}}$ | 175.66% | 115.81% | 78.86% | 63.73% |
| | ■ = $[ZT]_{int}$ | | 52.62% | 39.79% | 31.57% | 27.72% |
| | ■ = $Z_{int} T_{mid}$ | | 47.58% | 34.83% | 26.80% | 23.53% |
| | ■ = $Z_{eng} T_{mid}$ | | 8.09% | 9.19% | 11.47% | 13.33% |
| | ■ = $Z_0 T_{mid}$ (this work, **CSM**) | | 4.57% | 7.27% | 10.60% | 12.86% |
| Generic-Eng-ZT[16] using $Z_{eng}, W_T, W_J$ | | $\eta_c \dfrac{\sqrt{1+(ZT)_{eng}\alpha_1 \eta_c^{-1}}-1}{\sqrt{1+(ZT)_{eng}\alpha_1 \eta_c^{-1}}+\alpha_2}$ | 7.80% | 7.43% | 7.72% | 8.64% |
| 3DOFs[7] | (Z00) | $\dfrac{\Delta T}{T_h'} \cdot \dfrac{\sqrt{1+Z_{gen}T_{mid}'}-1}{\sqrt{1+Z_{gen}T_{mid}'}+\dfrac{T_c'}{T_h'}}$ | 4.13% | 6.85% | 10.56% | 12.66% |
| | ($Z\tau 0$) | | 0.87% | 1.24% | 1.47% | 1.50% |
| | ($Z\tau\beta$) | | 0.06% | 0.15% | 0.31% | 0.51% |

# Figure captions

**Figure 1.** Geometry of the thermoelectric generator module having length of $L$ area of $A$ under temperature difference of $\Delta T = T_h - T_c$. The single leg in gray is in contact with the hot side in red and the cold side in blue. Inside the TGM, the current flows from the hot to the cold side, and then flow through the load resistance $R_L$.

**Figure 2.** Numerical optimal efficiencies for materials with linearly temperature-dependent thermoelectric properties versus the slope of the Seebeck coefficient are calculated under $T_h = 900\ K$ and $T_c = 300\ K$. Each color represents a different CSA ZT value ($Z_0 T_{mid}$).

**Figure 3.** Relative error in formula optimal efficiency calculations for generated thermoelectric properties having $Z_0 T_{mid}$ value of (a) 1.0 and (b) 3.0.

**Figure 4.** Variation of device parameters and heat currents for TGMs with various TEP curve sets for working temperatures of $T_h = 900\ K$ and $T_c = 300\ K$. (a) for $Z_0 T_{mid} = 1$ with $\frac{\Delta \alpha}{2\alpha_0} = 0.4$, $\frac{\Delta \rho}{2\rho_0} = -0.32$, $\frac{\Delta \kappa}{2\kappa_0} = -0.32$; (b) for $Z_0 T_{mid} = 2$ with $\frac{\Delta \alpha}{2\alpha_0} = 0.2$, $\frac{\Delta \rho}{2\rho_0} = 0.16$, $\frac{\Delta \kappa}{2\kappa_0} = -0.16$; (c) for $Z_0 T_{mid} = 3$ with $\frac{\Delta \alpha}{2\alpha_0} = 0.6$, $\frac{\Delta \rho}{2\rho_0} = -0.32$, $\frac{\Delta \kappa}{2\kappa_0} = 0.8$, where $\Delta$ represents the difference between hot and cold sides. Relative $R$ and $K$ with respect to the open-circuit values $R_0$ and $K_0$, respectively, and relative heat current at hot side $Q_h$ with respect to the $Q_h^{(CSM)}$ are plotted versus relative current $i$. Th relative current ranges from 0 to the maximum current of a given TGM where load resistance is zero.

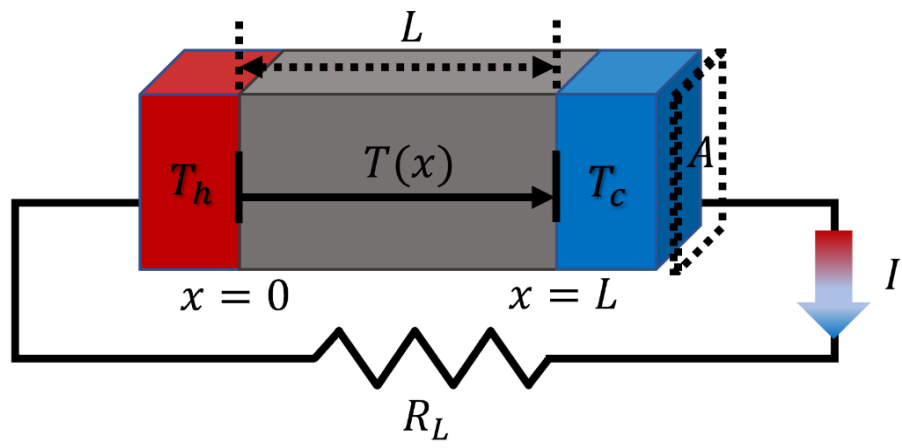

**Figure 1**

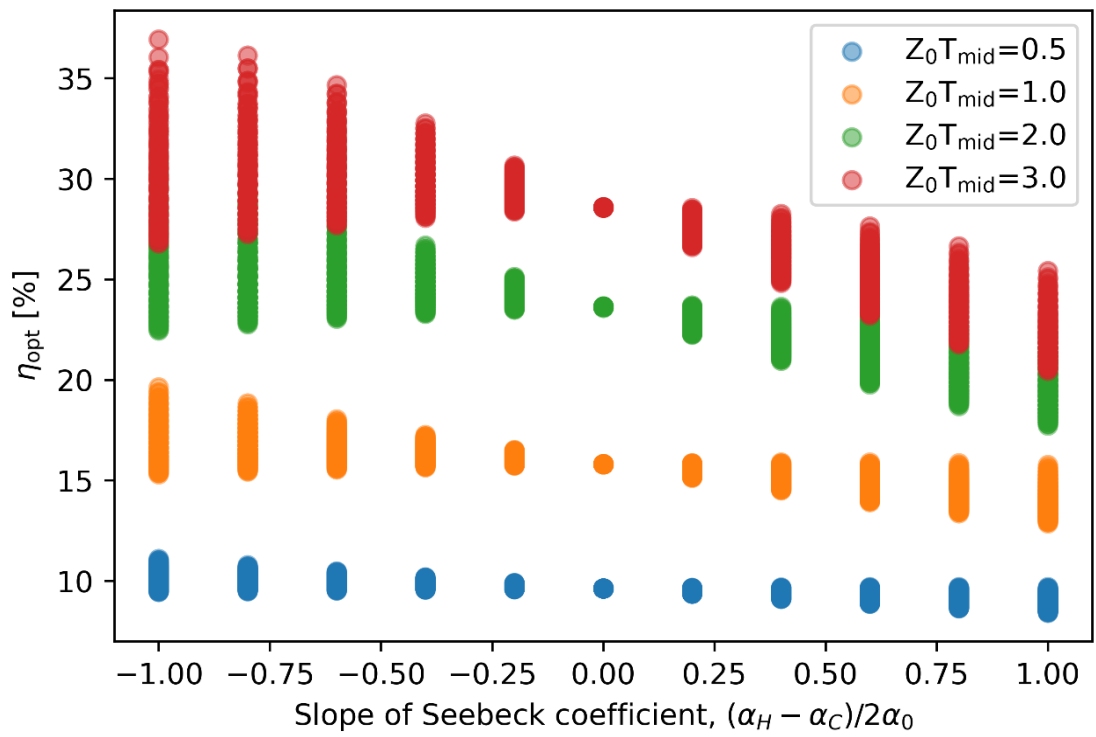

**Figure 2**

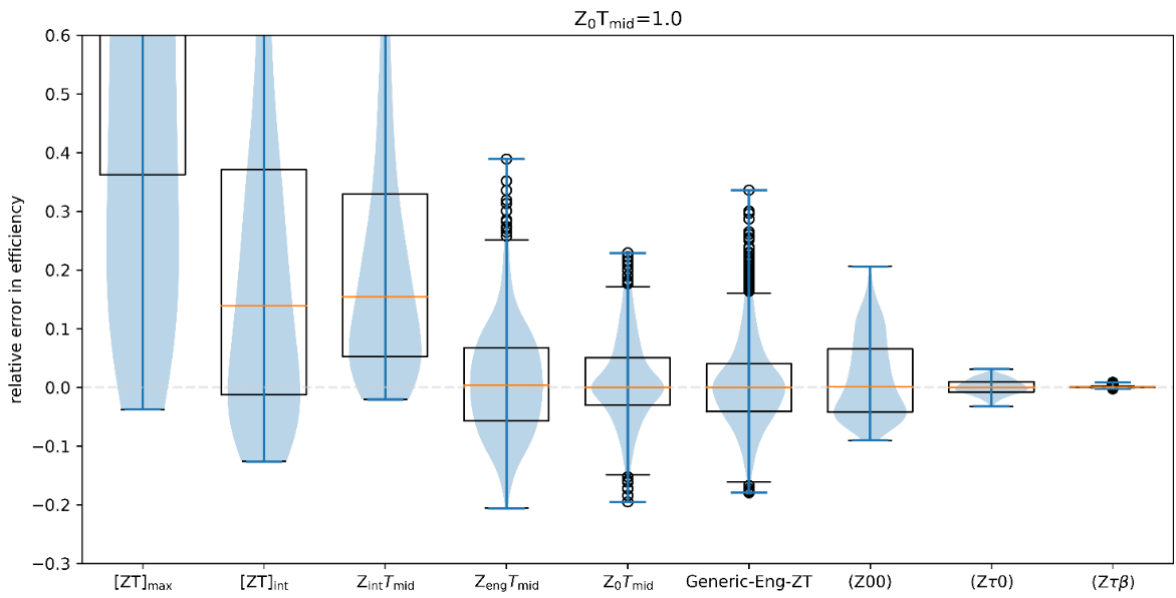

Figure 3(a)

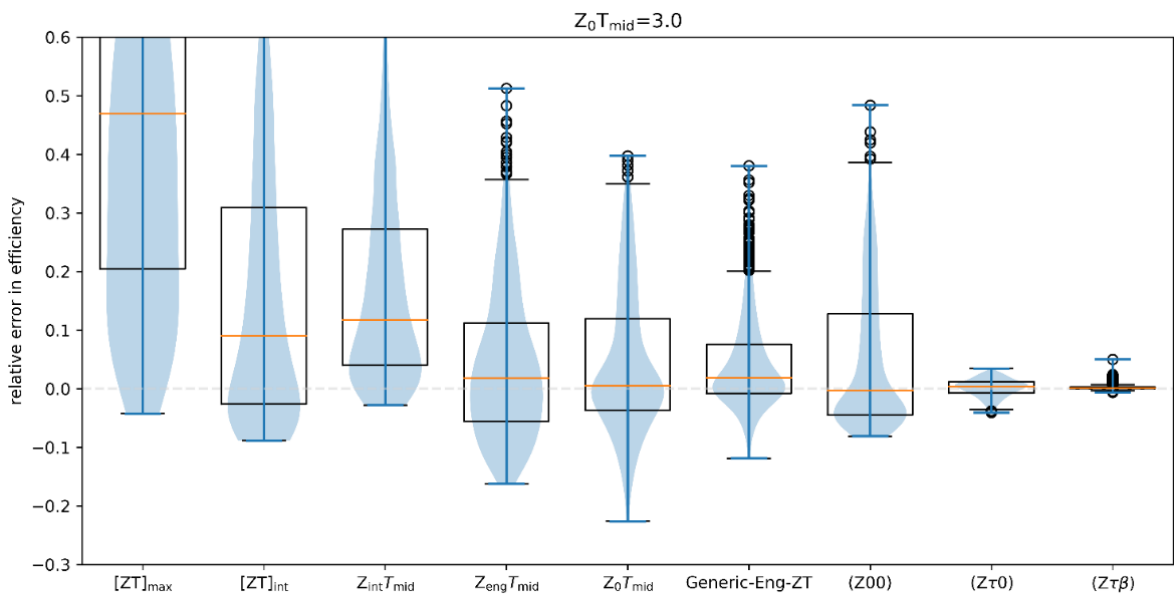

Figure 3(b)

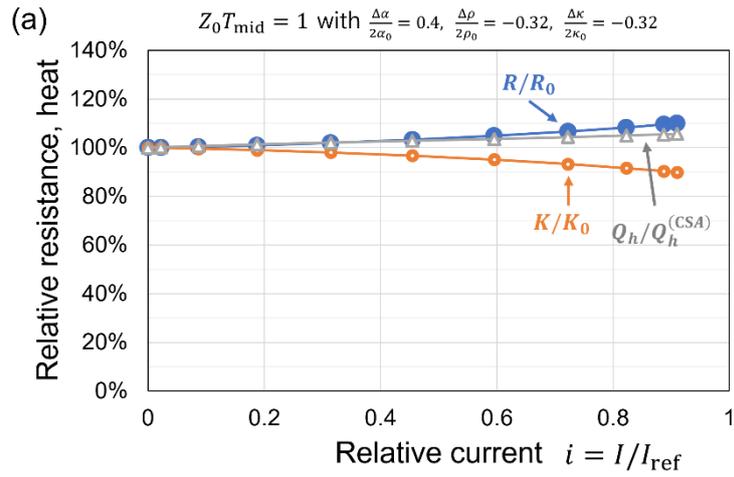

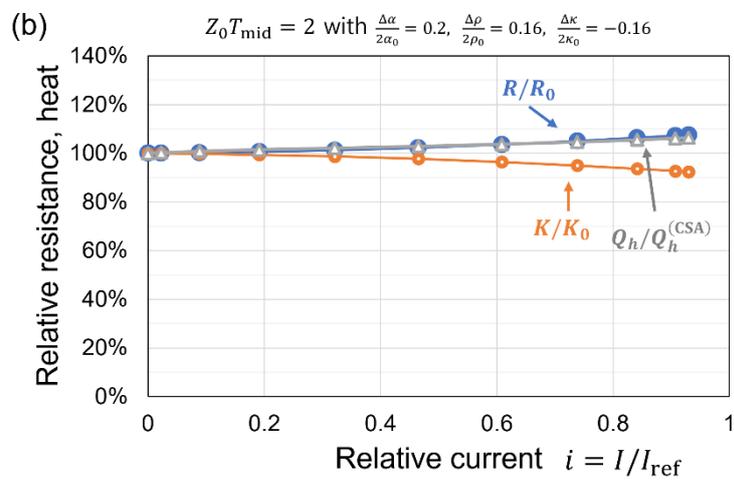

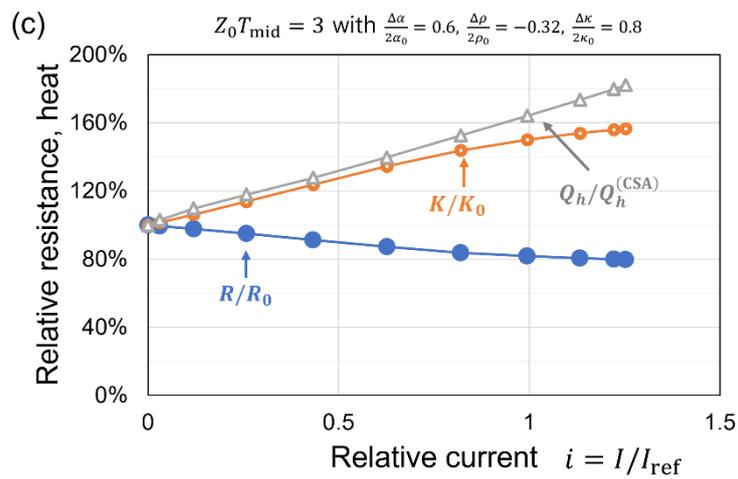

**Figure 4**